\documentclass[graybox]{svmult}

\usepackage{mathptmx} % selects Times Roman as basic font
\usepackage{helvet} % selects Helvetica as sans-serif font
\usepackage{courier} % selects Courier as typewriter font
\usepackage{type1cm} % activate if the above 3 fonts are
\usepackage{makeidx} % allows index generation
\usepackage{graphicx} % standard LaTeX graphics tool
\usepackage{multicol} % used for the two-column index
\usepackage[bottom]{footmisc}% places footnotes at page bottom

\makeindex % used for the subject index
\begin{document}

\title*{Determination of the physical characteristics of the variable 
stars in the direction of the open cluster NGC 6811 through $uvby\beta$ photoelectric photometry$^{*}$}
\titlerunning{Variable stars in NGC 6011}
\author{J.H. Pe\~na$^{1}$, L. Fox-Machado$^{2}$, H. Garc\'{\i}a$^{3}$, A. Renter\'{\i}a$^{1,4}$, E. Romero$^{1,4}$, S. Skinner$^{4,5}$, A. Espinosa$^{4}$}
\authorrunning{J.H. Pe\~na, et al.}
\institute{$^{*}$Based on observations at SPM, Mexico\\ $^{1}$Instituto de Astronom\'{\i}a, UNAM, M\'exico, \email{jhpena@astroscu.unam.mx}
\and $^{2}$Observatorio Astron\'omico Nacional, UNAM, Ensenada, B.C. M\'exico \and $^{3}$Observatorio Astron\'omico Nacional 
de la UNAN-Managua \and $^{4}$Facultad de Ciencias, UNAM, M\'exico \and $^{5}$Facultad de F\'{\i}sica, Universidad de Panam\'a}

\maketitle
% Please use both starred abstract and non-starred abstract.

\section{Background} The study of open clusters and their short period variable stars is fundamental in stellar evolution. Because the cluster members are formed in almost the same physical conditions, they share similar stellar properties such age and chemical composition. The assumption of common age, metallicity and distance impose strong constraints when modeling an ensemble of short period pulsators belonging to open clusters (e.g. Fox Machado et al., 2006).
Very recently, Luo et al. (2009) carried out a search for variable stars in the direction of NGC 6811 with CCD photometry in B and V bands. They detected a total of sixteen variable stars. Among these variables, twelve were catalogued as $\delta$ {\it Scuti} stars, while no variability type was assigned to the remaining stars. They claim that the twelve $\delta$ {\it Scuti} stars are all very likely members of the cluster which makes this cluster an interesting target for asteroseis-mological studies. Moreover, NGC 6811 has been selected as a asteroseismic target of the Kepler space mission (Borucki et al. 1997). Therefore, deriving accurate physical parameters for the pulsating star members is very important.

\section{Observations} These were all taken at the Observatorio Astronomico Nacional, Mexico in two different seasons (2009 and 2010). The 1.5 m telescope to which a spectrophotometer was attached was utilized at all times.

\section{Methodology and Analysis}  The evaluation of the reddening was done by first establishing to which spectral class the stars belonged: early (B and early A) or late (late A and F stars) types; the later class stars (later than G) were not considered. In order to determine the spectral type of each star, the location of the stars on the $[m_1] - [c_1]$ diagram (Golay, 1974, see below) was employed as a primary criteria. The reddening determination was obtained from the spectral types through Str\"omgren photometry. The application of the calibrations developed for each spectral type (Shobbrook, 1984 for O and early A types, and Nissen, 1988 for late A and F stars) were considered. The membership was determined from the Distance Modulus or distance histograms. We can establish that NGC 6811 has a distinctive accumulation of thirty-seven stars at a distance modulus of $10.5 \pm 1.0$~mag, Age is fixed for the clusters once we measured the hottest, and hence, the brightest stars . We have utilized the $(b-y)$ vs. $c_0$ diagrams of Lester, Gray \& Kurucz (1986) which allow the determination of the temperatures with an accuracy of a few hundreds of degrees

\section{Variable stars}  We carried out some very short span observations in differential photometric mode. The variables we considered were chosen due to their nearness and were, in the notation of Luo et al. (2009): V2, V4, V11 and V14 with W5 and W99 as reference and check stars. Although the time span we observed was too short to detect long period variation, the only star which showed clear variation was V4, with two clearly discernible peaks and of relatively large amplitude of variation $0.188$~mag and a period of $0.025$~d

\section{Discussion}  We found that the cluster is farther, its extinction is less and it is younger than previously assumed. The goodness of our method has been previously tested, as in the case of the open cluster Alpha Per (Pe\~na \& Sareyan, 2006) against several sources which consider proper motion studies as well as results from Hipparcos and Tycho data basis. Hence, we feel that our results throw new light regarding membership to this cluster.

Memberships are determined for V1, V2, V3, V5, V6, V10, V11, V13 and V16. Non-membership for V4, V12, V14, and V15 and we were unable to determine membership for V18 mainly because it does not belong to the spectral classes B, A or F and belongs to a latter spectral type which makes it impossible for it to be a $\delta$ {\it Scuti} type variable. From the location of these variables in the theoretical grids of LGK86 (see corresponding Figures) we determine their temperatures.

\begin{acknowledgement}
We would like thank the staff of the OAN for their assistance through the observations. This paper was partially supported by Papiit IN 114309. J. Orta and J. Miller did typing and proofreading, respectively; C. Guzm\'an, A. D\'{\i}az \& F. Ruiz assisted us at the computing.

\end{acknowledgement}

\end{document}